\documentclass[11pt,a4paper]{article}

\renewcommand{\author}{Emil Khalisi}
\newcommand{\paperlabel}{Gripp \& Khalisi (2021)}
\newcommand{\titel}{Appulses of Jupiter and Saturn}
\newcommand{\version}{Version 2.14}
\renewcommand{\date}{\today}

\usepackage{hyperref}
\hypersetup{pdfauthor={Emil Khalisi}}
\hypersetup{pdftitle={\titel }}
\hypersetup{pdfsubject={Online publication, \version\ of \date\ }}
\hypersetup{pdfkeywords={Jupiter, Saturn, Great Conjunction, Trigons}}
\hypersetup{colorlinks=true,citecolor=blue}

\usepackage[T1]{fontenc}        
\usepackage{mathptmx}           
\usepackage{pifont}             
\usepackage[scaled=0.92]{helvet} 

\usepackage[british]{babel}     
\usepackage{amsmath}            
\usepackage{microtype}          
\usepackage{calc}
\usepackage[a4paper]{geometry}  
\usepackage{scrextend}           
\changefontsizes{10pt}

\usepackage{titlesec}
\titleformat*{\section}{\large\bfseries}
\titleformat*{\subsection}{\normalsize\bfseries}

\usepackage{graphicx}           
\usepackage{float}              
\usepackage{caption}            
\usepackage{fancyhdr}
\renewcommand{\headrulewidth}{0.4pt}
\usepackage{colortbl}             
\definecolor{grey20}{RGB}{208,208,208}

\usepackage{url}
\usepackage[numbers]{natbib}      
\begin{document}


\fancyhead{}
\fancyhead[LO]{%
   \footnotesize \textsc{In original form published in:}\\
   {\footnotesize Sternzeit 46, No.\ 1+2 / 2021 (ISSN: 0721-8168)}
}
\fancyhead[RO]{
   \footnotesize {\tt arXiv:(side label) [physics.pop-ph]}\\
   \footnotesize {Date: 6th May 2021}%
}
\fancyfoot[C]{\thepage}

\renewcommand{\abstractname}{}

\twocolumn[
\begin{@twocolumnfalse}

\section*{\centerline{\LARGE \titel }}

\begin{center}
{Joachim Gripp, \author \\}
\textit{Sternzeit e.V., Kiel and Heidelberg, Germany}\\
\textit{e-mail:} $\;\;$ \texttt{gripp}$\;$ or $\;$
       \texttt{khalisi$\;$ \dots [at]sternzeit-online[dot]de}
%
\end{center}


\vspace{-\baselineskip}
\begin{abstract}
\changefontsizes{10pt}
\noindent
\textbf{Abstract.}
The latest conjunction of Jupiter and Saturn occurred at an optical
distance of 6 arc minutes on 21 December 2020.
We re-analysed all encounters of these two planets between -1000
and +3000 CE, as the extraordinary ones ($<10^{\prime}$) take place
near the line of nodes every 400 years.
An occultation of their discs did not and will not happen within
the historical time span of $\pm$5,000 years around now.
When viewed from Neptune though, there will be an occultation in
2046.

\vspace{\baselineskip}
\noindent
\textbf{Keywords:}
Jupiter-Saturn conjunction,
Appulse,
Trigon,
Occultation.
\end{abstract}

\centerline{\rule{0.8\textwidth}{0.4pt}}
\vspace{2\baselineskip}

\end{@twocolumnfalse}
]


\section*{Introduction}

The slowest naked-eye planets Jupiter and Saturn made an impressive
encounter in December 2020.
Their approaches have been termed ``Great Conjunctions'' in former
times and they happen regularly every $\approx$20 years.
Before the discovery of the outer ice giants these classical planets
rendered the longest known cycle.
The separation at the instant of conjunction varies up to 1 degree
of arc, but the latest meeting was particularly tight since the
planets stood closer than at any other occasion for as long as 400
years.


\section*{Periodic Overtaking}

The precise average of the conjunction interval is 19.88 years and
represents the synodic period, $T_{\rm syn}$, when two moving bodies
apparently meet as seen from a third body (Earth):
\begin{equation*}
\frac{1}{T_{\rm syn}} = \frac{1}{T_{\rm Jup}} - \frac{1}{T_{\rm Sat}} \; ,
%
%
\end{equation*}
with $T_{\rm Jup}$ and $T_{\rm Sat}$ being the heliocentric orbital
periods of Jupiter and Saturn, respectively.
Because of the ellipticity of the orbits, $T_{\rm syn}$ fluctuates
between 18.85 and 21.07 years.

A threefold encounter within one year, called ``triple conjunction'',
can occur during the retrograde motion of the planets.
It happens at irregular times when two conditions are fulfilled:
(a) the times of their particular opposition will lie within
$\pm$1.7 days, and
(b) the viewing angle from Earth be less than 30$^{\circ}$ with regard
to the heliocentric position of the pair:
\begin{equation*}
  L - E \; < \; 30^{\circ} \; ,
\end{equation*}
wherein $L$ and $E$ are the heliocentric longitude of the pair and
of the Earth, respectively.

Triple conjunctions make up a subset of the 20-year meetings.
They occur once in 140 years on average, however, the interval may
be as little as 40 years or as much as 378 years \cite{weitzel_1945}.
There is no pattern in the triples, and
the reason is due to Earth's orbit:
while Jupiter and Saturn are locked in a 5:2-mean motion resonance,
the Earth does not join in.
For very long periods there could be some periodicity, however,
secular effects destroy a cycle, e.g.\ rotation of the apsides and
changes in eccentricity such that we are left with some kind of
``semi-periodicity''.


\section*{Close Encounters}

Most pass-bys of Jupiter and Saturn are not very spectacular,
their separation would exceed 1 degree of arc quite often.
We did some little exercise using the simulation software packages
\textit{Guide} v9.0 and \textit{Cartes du Ciel} v4.1 to search for
especially tight conjunctions, so-called ``appulses''.
Table \ref{tab:closeconjunctions} lists incidents less than
10$^{\prime}$ in the years 1000 BCE to CE 3000.
An eye-striking feature is that many events appear in doublets of
60 years.

\begin{table}[t]
\caption{Appulses of Jupiter and Saturn between 1000 BCE and CE 3000
    with a separation $<10^{\prime}$.}
\label{tab:closeconjunctions}
%
\vspace{0.5ex}
\centering
\begin{tabular}{rl|rc}
\hline
\rowcolor{grey20}
   \multicolumn{2}{c|}{\cellcolor{grey20}Date} &  Separ. & Elong.\\
\hline
%
  -998 & May 30 & 9$^{\prime}$ 19$^{\prime\prime}$ & 39 E \\
  -939 & Sep 04 & 3$^{\prime}$ 29$^{\prime\prime}$ & 42 W \\
  -482 & Mar 06 & 6$^{\prime}$ 23$^{\prime\prime}$ & 55 W \\
  -423 & Dec 28 & 1$^{\prime}$ 29$^{\prime\prime}$ & 17 E \\ 
   -85 & Aug 11 & 3$^{\prime}$ 44$^{\prime\prime}$ & 20 W \\
   372 & Mar 06 & 1$^{\prime}$ 52$^{\prime\prime}$ & 53 W \\
   431 & Dec 31 & 6$^{\prime}$ 16$^{\prime\prime}$ & 17 E \\
   709 & Sep 13 & 8$^{\prime}$ 22$^{\prime\prime}$ & 61 W \\ 
   769 & Jul 23 & 4$^{\prime}$ 17$^{\prime\prime}$ &  2,4 W \\
  1166 & Dec 11 & 9$^{\prime}$ 48$^{\prime\prime}$ & 25 E \\
  1226 & Mar 05 & 2$^{\prime}$ 09$^{\prime\prime}$ & 49 W \\
  1563 & Aug 25 & 6$^{\prime}$ 47$^{\prime\prime}$ & 42 W \\
  1623 & Jul 16 & 5$^{\prime}$ 10$^{\prime\prime}$ & 13 E \\
  2020 & Dec 21 & 6$^{\prime}$ 17$^{\prime\prime}$ & 30 E \\ 
  2080 & Mar 15 & 6$^{\prime}$ 19$^{\prime\prime}$ & 44 W \\ 
  2417 & Aug 24 & 5$^{\prime}$ 25$^{\prime\prime}$ & 26 W \\ 
  2477 & Jul 06 & 6$^{\prime}$ 20$^{\prime\prime}$ & 27 W \\ 
  2815 & Feb 23 & 9$^{\prime}$ 58$^{\prime\prime}$ & 30 W \\ 
  2874 & Dec 25 & 2$^{\prime}$ 21$^{\prime\prime}$ & 36 E \\
  2934 & Mar 19 & 9$^{\prime}$ 43$^{\prime\prime}$ & 38 W \\
\hline
\end{tabular}
\end{table}

\begin{figure}[t]
\centering
\includegraphics[width=0.85\linewidth]{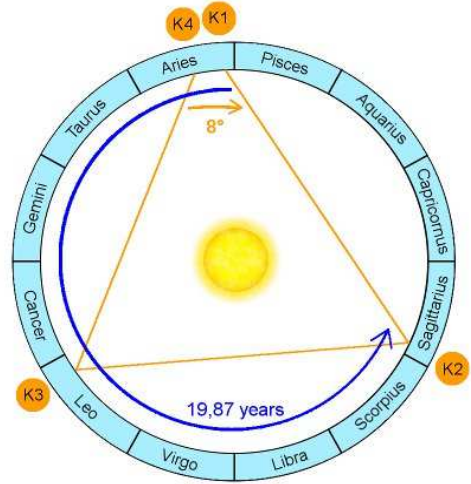}
\caption{Displacement of a Trigon after 60 years.}
\label{fig:trigone}
\end{figure}

The explanation for the 60-year gap is based on ``Trigons'':
after 3$\times$20 years the conjunction occurs in the same
constellation, just 8$^{\circ}$ apart from the former position
(Fig.\ \ref{fig:trigone}).
The missing piece makes the Trigon rotate slowly in front of the
stellar background in 794 year's time, often rounded up to 800
years.
During the Middle Ages this period inspired astrologers to look
for profane meanings in world history.
The source of such ideas may have been Hindu (3rd to 5th
century CE) transmitted via Sassanid Persia to Europe
\cite{etz_2000}.

Those extraordinary close conjunctions, that we were searching for,
take place when both planets meet close to their line of nodes.
The orbital planes are inclined to each other at an angle of
1.2$^{\circ}$, and the line of intersection points at Capricorn and
Cancer, respectively, in our era.
The nodes move very slowly retrograde, but remain for 4,000 years
inside a constellation.

\begin{figure}[t]
\includegraphics[width=\linewidth]{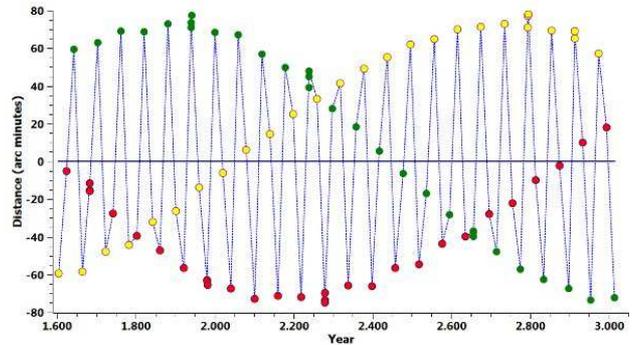}
\caption{Separation of Jupiter and Saturn at the instant of their
    conjunction.
    Negative values indicate Jupiter passing southward.
    The three sinusoidal curves represent one leg of the Trigon.}
\label{fig:conjcycles}
\end{figure}

The latest conjunction of 2020 was observed in Capricorn, and the
next of comparable apparent distance, in 2080, will come about in
Capricorn again.
However, Jupiter will pass northward of Saturn since both planets
will have already left behind the ascending node
(Fig.\ \ref{fig:conjcycles}).
The 8$^{\circ}$-shift of the Trigon leads to the effect that
appulses are precluded for the subsequent centuries.
Only when one of its legs approaches a node, there will be tight
encounter:
thereafter in 2417 and 2477 at the opposite (descending) node.

%
\renewcommand{\headrulewidth}{0pt}
\fancyhead{}
\fancyhead[C]{\footnotesize \itshape \paperlabel : \titel}


\section*{Planetary Occultations}

A brief remark in an article by Johan Stein \cite{stein_1949}
informs us about an observation by the astrologer Francesco Giuntini
(1523--1590):
Jupiter almost occulted Saturn on 24 August 1563.
In fact, the planets passed by each other similarly close as
in 2020.
In the pre-telescopic era the shining objects seemed merging.
The resolution of the normal eyesight is at 1$^{\prime}$, but this
turns out a matter of contrast and individual training of the
observer.

Mutual occultations of any two planets in the Solar System do exist
within the time span of our historic era during $\pm$3000 years,
but there is none for Jupiter and Saturn.
This oddity was already noticed by Salvo de Meis \cite{demeis_1993}.
We made an effort to look a bit further for an overlap of their
planetary discs.
A promising case is found for 16 February 7541 using the ephemeris
of the DE431 theory of planetary motions.
Near misses are going to happen in 4523 and 6687.

However, extrapolations of this kind can hardly be verified, as they
put astronomical calculations to extremes.
All motions in the Solar System are extrapolated from our narrow
window of precise measurements.
The movements are prone to gravitational perturbances, and even the
slightest deviation might cause a delay.
Moreover, the diameters of the planets are slowly varying with time
by reason of secular variations of their orbits \cite{meeus_1970}.
For very distant times any prediction of an accuracy smaller than
the apparent diameters of the planets cannot be made,
as long as we lack secure fixpoints from observation to be utilised
in a wider baseline for extrapolation purposes.
The characteristics considered to be stable for millions of years
regard only the orbital parameters, not to be confused with the
ephemeris.
A comparison of various software deploying different integrators
would be desirable though.

\begin{figure}[t]
\centering
\includegraphics[width=\linewidth]{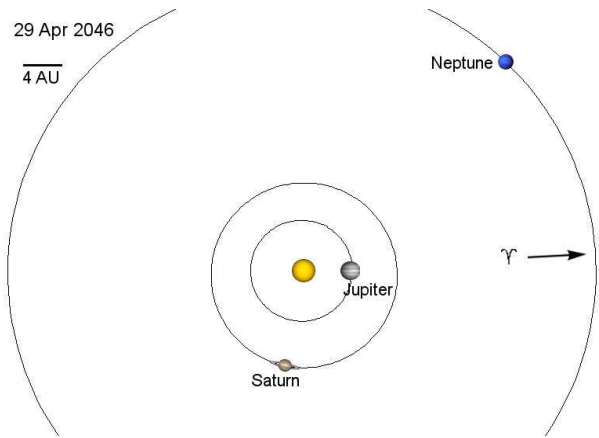}
\caption{Occultation of Saturn and Jupiter on 29 April 2046 viewed from Neptune.}
\label{fig:neptune}
\end{figure}

Still, there is a chance to spot a line-up of Jupiter and Saturn:
that special view emerges extraterrestrically from Neptune on
29 April 2046 (Fig.\ \ref{fig:neptune}).
Gazing from this furthermost planet, Saturn is on the superior arc
of its orbit in a distance of 38.7 AU from Neptune while Jupiter
is heading to its inferior conjunction with the sun being 26.7 AU
away.
The planets meet in an optical line 8$^{\circ}$ to the east of the
sun.
It will be the sole occultation event for these two giants visible
from any planet during the entire 21st century.

Besides the case of Jupiter and Saturn, there is also no occultation
for the remote Uranus and Neptune to be seen from Earth.
That seems intelligible as their conjunctions take place only at 170
years on average and their orbits are inclined by 10$^{\circ}$.
An overlap of their discs will probably not occur within 1 million
years.

In regard to the Jovian planets, it is actually Saturn who refuses
to meet its giant colleagues.
While Uranus and Neptune are hit by Jupiter once or twice during our
historical time, Saturn resists any such visit with the outer planets.
As viewed from Earth, it is only occulted by the inner planets.
The next occultation will be performed by Venus on 12 August 2243.



\section*{Acknowledgements}

This paper is a summary of two articles in the German journal for
astronomy \textit{Sternzeit 46}, issue no.\ 1 + 2/2021, p18--23 and
p69--71.


\vspace{\baselineskip}

\bibliography{v213-biblio}

\end{document}